\documentclass[twocolumn,showpacs,preprintnumbers,amsmath,amssymb]{revtex4}
\usepackage{graphicx}
\usepackage{dcolumn}
\usepackage{bm}

\begin{document}

\def\ket#1{\langle#1\mid}
\def\bra#1{\mid#1\rangle}


\title{Local entanglement and quantum phase transition in the Hubbard model}

\author{Shi-Jian Gu$^{1,2}$}
\author{You-Quan Li$^2$} \author{Hai-Qing Lin$^1$}
\affiliation{$^1$Department of Physics, The Chinese University of
Hong Kong, Hong Kong, China} \affiliation{$^2$Zhejiang Institute
of Modern Physics, Zhejiang University, Hangzhou 310027, P. R.
China}

\begin{abstract}

The local entanglement $E_v$ of the one-dimensional Hubbard model
is studied on the basis of its Bethe-ansatz solution. The
relationship between the local entanglement and the on-site
Coulomb interaction $U$ is obtained. Our results show that $E_v$
is an even analytic function of $U$ at half-filling and it reaches
a maximum at the critical point $U=0$. The variation of the local
entanglement with the filling factor shows that the ground state
with maximal symmetry possesses maximal entanglement. The magnetic
field makes the local entanglement to decrease and approach to
zero at saturated magnetization. The on-site Coulomb interaction
always suppresses the local entanglement.

\end{abstract}

\maketitle

Quantum entanglement, as one of the most intriguing feature of
quantum theory, has been a subject of much study in recent years,
mostly because its nonlocal connotation\cite{ABinstein35} is
regarded as a valuable resource in quantum communication and
information processing \cite{SeeForExample,MANielsenb}. For the
application purpose, much of recent attentions have been focused
on entanglement relevant to realistic systems. For example,
several authors have investigated entanglement in spin systems
\cite{KMOConnor2001,XWang_PRA_64_012313,GLagmago2002,TJOsbornee,AOsterloh2002,GVidal2003}
as well as indistinguishable-particle systems
\cite{JSchliemann_PRB_63_085311,PZanardi_PRA_65_042101}. The work
of Osterloh {\it et al.}\cite{AOsterloh2002} and Osborne and
Nielsen\cite{TJOsbornee} on the XY model suggestively showing that
the entanglement of two neighboring sites displays a sharp peak
either near or at the critical point where quantum phase
transition undergoes. Investigating the critical entanglement
between a block of continuous spins and their supplemental parts
in a spin chain model, Vidal {\it et al.} \cite{GVidal2003}
pointed out its relation to the entropy in conformal field
theories. Recently, we studied entanglement and quantum phase
transition of the XXZ model \cite{SJGuup} and obtained its
dependences on the anisotropy parameter $\Delta$ and correlation
length $\xi$.

It is well known that the Hubbard model is a typical model
describing correlated Fermion systems. It plays a crucial role for
understanding many physical phenomena in condensed matter physics,
such as magnetic ordering, Mott-insulator transition, and
superconductivity, etc. Therefore, the study of the entanglement
in the Hubbard model not only provides possible clues for
experimental realization, but also sheds new light on the
understanding of quantum many-body systems.

In this Letter, we study the local entanglement of one-dimensional
Hubbard model\cite{JHubbard63}. We obtain the dependence of the
local entanglement on Coulomb interaction, particle number and
magnetization respectively. We find that the ground state with
maximal symmetry possesses the maximal local entanglement, and the
Coulomb interaction always suppresses the local entanglement. To
the best of our knowledge, no one has investigated the
entanglement in the interacting many-fermion systems. Our results,
which are based on the exact solution of the Hubbard model, will
be inspirable for people to explore quantum entanglement and phase
transition via nonperturbative approach for other interacting
many-fermion systems. An important observation of our studies is
that the local entanglement is a smoothly continuous function at
the quantum phase transition (Mott-transition) point, which
clarifies that the singularity of the entanglement at quantum
phase transition point is not an universal behavior as suggested
by previous studies.

The Hamiltonian of the one-dimensional Hubbard model reads
\begin{eqnarray}\label{eq:Hamiltonian}
H=-\sum_{\sigma,j}(c^\dagger_{j\sigma}c_{j+1 \sigma}+c_{j+1
\sigma}^\dagger c_{j \sigma})+U \sum_j n_{j
 \uparrow}n_{j \downarrow}
\end{eqnarray}
where $\sigma=\uparrow,\downarrow;\,j=1,\dots, L$, $c^\dagger_{j
\sigma}$ and $c_{j \sigma}$ create and annihilate respectively an
electron of spin $\sigma$ at site $j$. There are  four possible
local states, $|0\rangle_j,\,|\uparrow\rangle_j,\,
|\downarrow\rangle_j,\,|\uparrow\downarrow\rangle_j$ denoted by $
|\nu\rangle_j,\, \nu=1,2,3,4 $.
 The Hilbert space of
$L$-site system is of $4^L$  dimensional, and
$|\nu_1,\,\nu_2\,\cdots\nu_L\rangle= \prod_{j=1}^L|\nu_j
\rangle_j$ are its natural basis vectors. Therefore any state in
such a system can be expressed as a superposition of the above
basis vectors. We consider the ground state $|\Psi\rangle$ of the
model. The local density matrix is a reduced density matrix
$\rho_j={\rm Tr}_j|\Psi\rangle\langle\Psi|$, where ${\rm Tr}_j$
stands for tracing over all sites except the $j$th site. This
enables us to study the correlations between the two parts of the
Hilbert subspace defined on
$\mathbb{C}^4\otimes(\mathbb{C}^4)^{L-1}$. Accordingly, the von
Neumann entropy $E_v$ calculated from the reduced density matrix
$\rho_j$ is employed to measure the entanglement of states on the
$j$th site with that on the remaining $L-1$ sites. It was called
the local entanglement\cite{PZanardi_PRA_65_042101} for it
exhibits the correlations between a local state and the other part
of the system.

The Hamiltonian (\ref{eq:Hamiltonian}) possesses U(1)$\times$SU(2)
symmetry, i.e., it is invariant under the gauge transformation
$c_{j,\sigma}\rightarrow e^{i\theta}c_{j,\sigma}$ and spin
rotation $c_{j,\sigma}\rightarrow U_{\sigma\delta}c_{j,\delta}$,
which manifest the charge and spin conservation. Thus, solutions
of (\ref{eq:Hamiltonian}) should simultaneously be the eigenstates
of particle number $N=\sum_j ( n_{j,\uparrow}+n_{j,\downarrow})$
and the $z$-component of total spin $S^z=\sum_j(
n_{j,\uparrow}-n_{j,\downarrow})$. This implies the absence of
coherent superposition between configurations with different
eigenvalues of $n_{j,\uparrow}+n_{j,\downarrow}$ and
$n_{j,\uparrow}-n_{j,\downarrow}$.
The density matrix $\rho_j$ therefore has the following diagonal
form
\begin{equation}
\rho_j = z\bra{0}\ket{0} + u^+\bra{\uparrow}\ket{\uparrow}
         + u^- \bra{\downarrow}\ket{\downarrow}
          + w \bra{\uparrow\downarrow}\ket{\uparrow\downarrow}
\end{equation}
Parameters in the density matrix $\rho_j$ represent local
populations of four states, i.e.,
\begin{eqnarray}
w&=&\langle n_{j\uparrow}n_{j\downarrow}\rangle
 = {\rm Tr}(n_{j \uparrow}n_{j \downarrow}\rho_j), \nonumber\\
  u^+&=&\langle n_\uparrow\rangle - w,  \;\;
   u^-=\langle n_\downarrow\rangle - w, \nonumber\\
    z&=&1 - u^+ - u^- -w
    = 1 - \langle n_\uparrow\rangle - \langle n_\downarrow\rangle + w
\end{eqnarray}
where $\langle n_\uparrow\rangle$ and $\langle
n_\downarrow\rangle$ are electron density with spin up and spin
down respectively. Thus the key point in discussing the local
entanglement is to study the density distribution of the four
local states.

As the Hamiltonian is invariant under translation, the local
density matrix $\rho_j$ is site independent. Consequently, the von
Neumann entropy (or the local entanglement which we call
hereafter) of this local reduced density matrix is
\begin{eqnarray}
E_v=-z\log_2z-u^+\log_2u^+ -u^-\log_2u^- -w\log_2w \nonumber
\end{eqnarray}

The one-dimensional Hubbard model has been solved exactly by the
Bethe-ansatz method \cite{EHLieb68,MTakahashib}. The energy
spectra is given by $E=-2\sum_{j=1}^N\cos k_j$, for $U>0$ and
$N\leq L$. The charge and spin rapidities $\{k_j,\lambda_a\}$
fulfill the following transcendental equations:
\begin{eqnarray}\label{eq:BAE}
&&2\pi I_j=k_j L-\sum_{a=1}^M\theta_1(\lambda_a-\sin
k_j),\nonumber
\\
&&2\pi J_a=\sum_{j=1}^N\theta_1(\lambda_a-\sin k_j)
 -\sum_{b=1}^M\theta_2(\lambda_a-\lambda_b),
\end{eqnarray}
where $\theta_n(k)=2\tan^{-1}(4k/nU)$, $M$ is the number of
electrons with down spins, $I_j$ and $J_a$ play the role of
quantum numbers. The ground state is a spin singlet, whose
solution consists of real $k$s and $\lambda$s, which is solved by
choosing the quantum number configuration as successive integers
or half-odd-integers  symmetrically arranged around zero. For
cases of either $U<0$ or $N>L$ the energy spectra can be obtained
by a particle-hole transformation\cite{EHLieb68}
\begin{eqnarray}\label{eq:nUtran}
&&E(N-M, M; U)\nonumber \\
&&=(N-M)U + E(N-M, L-M; -U), \nonumber
\\
&&=-(L-N)U+E(L-N+M, L-M; U)
\end{eqnarray}

Now we are in the position to investigate the relationship between
the local entanglement and the on-site Coulomb coupling $U$, the
chemical potential, and the external magnetic field.

{\it Local entanglement at half-filling and quantum phase
transition driven by the on-site Coulomb coupling:}
>From the Lieb-Wu solution of the ground state at half-filling \cite{EHLieb68},
we obtain the density of double occupancy $w$ by taking the
derivative of $E_0(U)/L$ with respect to $U$ according to the
Hellman-Feynman theorem,
\begin{eqnarray}\label{eq:doublw}
w=\int_0^\infty\frac{J_0(\omega)J_1(\omega)d\omega}{1+\cosh(U\omega/2)}.
\end{eqnarray}
where $J_0(\omega)$ and $J_1(\omega)$ are zeroth and first order
Bessel functions. The ground state is a spin singlet which implies
$\langle n_\uparrow\rangle=\langle n_\downarrow\rangle=1/2$, hence
$u^+$ and $u^-$ are both simply $1/2-w$. The local entanglement is
then
\begin{eqnarray}
E_v=-2w\log_2w-2\left(1/2-w\right) \log_2\left(1/2-w\right).
\label{eq:localevhalffilling}
\end{eqnarray}
By making the use of Eq. (\ref{eq:nUtran}), one easily finds that
 $w(-U)= 1/2-w(U)$,
so the local entanglement is an even function of $U$
\begin{eqnarray}\label{eq:Evrelation}
E_v(-U)=E_v(U).
\end{eqnarray}

\begin{figure}
\includegraphics[width=7cm]{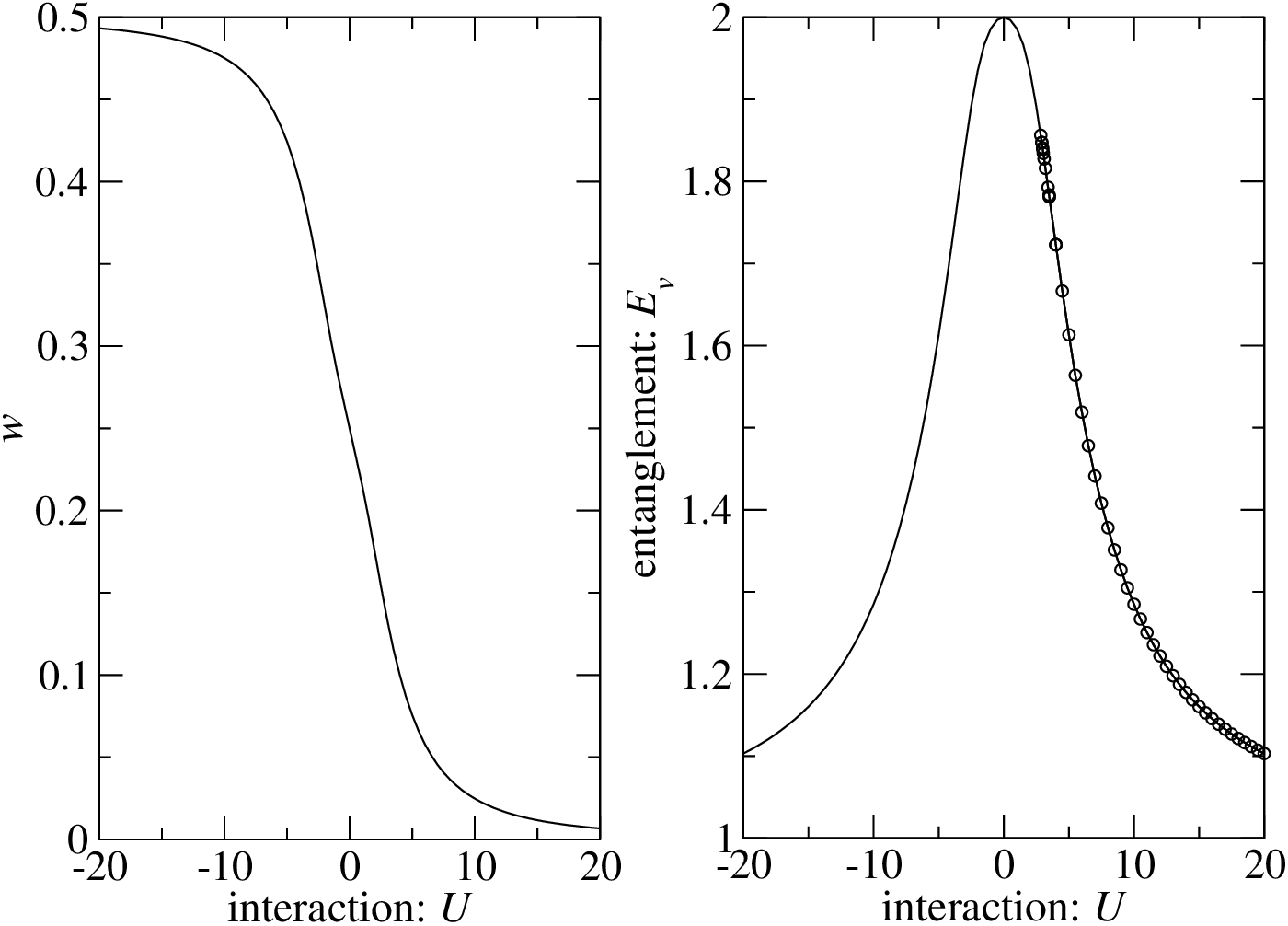}
\caption{\label{figure_ent} The density of double occupancy $w$
(left figure) and the local entanglement $E_v$ (right figure)
versus the on-site coupling $U$ . The solid lines are obtained
from Eq. (\ref{eq:doublw}) and (\ref{eq:localevhalffilling}),
while the open dots are obtained by solving the Eqs.
(\ref{eq:BAE}) of $N=70$ sites system numerically.}
\end{figure}

In the strong coupling limit, $U\rightarrow\infty$, the double
occupancy tends to zero and the sites are singly occupied.
The local density matrix $\rho_j$ has two nonvanishing values
$u^+=u^-=1/2$ that results in $E_v(\infty)=1$. For finite positive
$U$ the hopping process
$|\uparrow\rangle_j|\downarrow\rangle_{j+1}\rightarrow|
\uparrow\downarrow\rangle_j|0\rangle_{j+1}$ or
$|0\rangle_j|\uparrow\downarrow\rangle_{j+1}$
brings about the occurrence of double occupancies. The lower the
on-site repulsive coupling $U$ is, the larger the occurrence
likelihood will be. Consequently, decreasing $U$ will enhance the
local entanglement $E_v$. For negative $U$ case, however, double
occupancy is favored. If $U\rightarrow -\infty$, half of the total
sites are doubly occupied while the other half are empty, which
gives rise to $z=w=1/2$, $u^+=u^-=0$, and $E_v(-\infty)=1$. For
finite negative $U$, the hopping process
$|\uparrow\downarrow\rangle_j|0\rangle_{j+1}\rightarrow|
\uparrow\rangle_j|\downarrow\rangle_{j+1}$ or
$|\downarrow\rangle_j|\uparrow\rangle_{j+1}$
breaks up some of the double occupancies and suppresses the
magnitude of $w$ and $z$. This enhances the local entanglement,
$E_v$. As a result, the local entanglement increases when $|U|$
decreases. It is expected to reach a maximum value when $U$
approaches to zero from both positive and negative sides. We plot
the density of double occupancies $w$ and local entanglement $E_v$
as functions of $U$ in Fig. \ref{figure_ent}, obtained by
integrating Eq. (\ref{eq:doublw}). Data points in the figure are
obtained by solving Bethe-ansatz equation (\ref{eq:BAE})
numerically on a 70-site chain. The excellent agreement justifies
the validity of later calculations.

It is well known that the ground state of the one-dimensional
Hubbard model at half-filling is metallic for $U\leq 0$, and
insulating for $U>0$. The point $U=0$ separates metallic and
insulating phases. Our calculation shows that the local
entanglement reaches the maximum at the critical point $U=0$ where
quantum phase transition occurs. It is also worthwhile to notice
that the derivative of the local entanglement with respect to the
on-site coupling $U$ is negative in the insulating phase but
positive in the metallic phase.

Eq. (\ref{eq:doublw}) could be integrated by performing series
expansion in $U$ for the Bessel functions. In the large $U\gg 1$
region, to the third order in $1/U^2$, we have $ w=4\ln
2/U^2-27\zeta(3)/U^4+375\zeta(5)/U^6$ where $\zeta$ stands for the
Riemann zeta function. This shows that $w$ descends rapidly when
$U$ increases. Therefore the local entanglement yields the
following asymptotic behavior
\begin{eqnarray}
E_v=1+16\ln U/U^2+\cdots.
\end{eqnarray}
Whereas in the week coupling region $0<U\ll 1$, the density of
double occupancy becomes
$w=1/4-7\zeta(3)U/8\pi^3-93\zeta(5)U^3/2^9\pi^5$, which is
obtained by making the use of energy expansion with respect to $U$
\cite{ENEconomou79,WMetzner89}. Thus, the local entanglement near
the critical point is,
\begin{eqnarray}
E_v=2-\frac{1}{\ln 2}\left[\frac{7\zeta(3)
U}{2\pi^3}\right]^2+\cdots.
\end{eqnarray}
Clearly, $E_v$ is analytic in the neighborhood of the critical
point $U=0$.

\begin{figure}
\includegraphics[width=7cm]{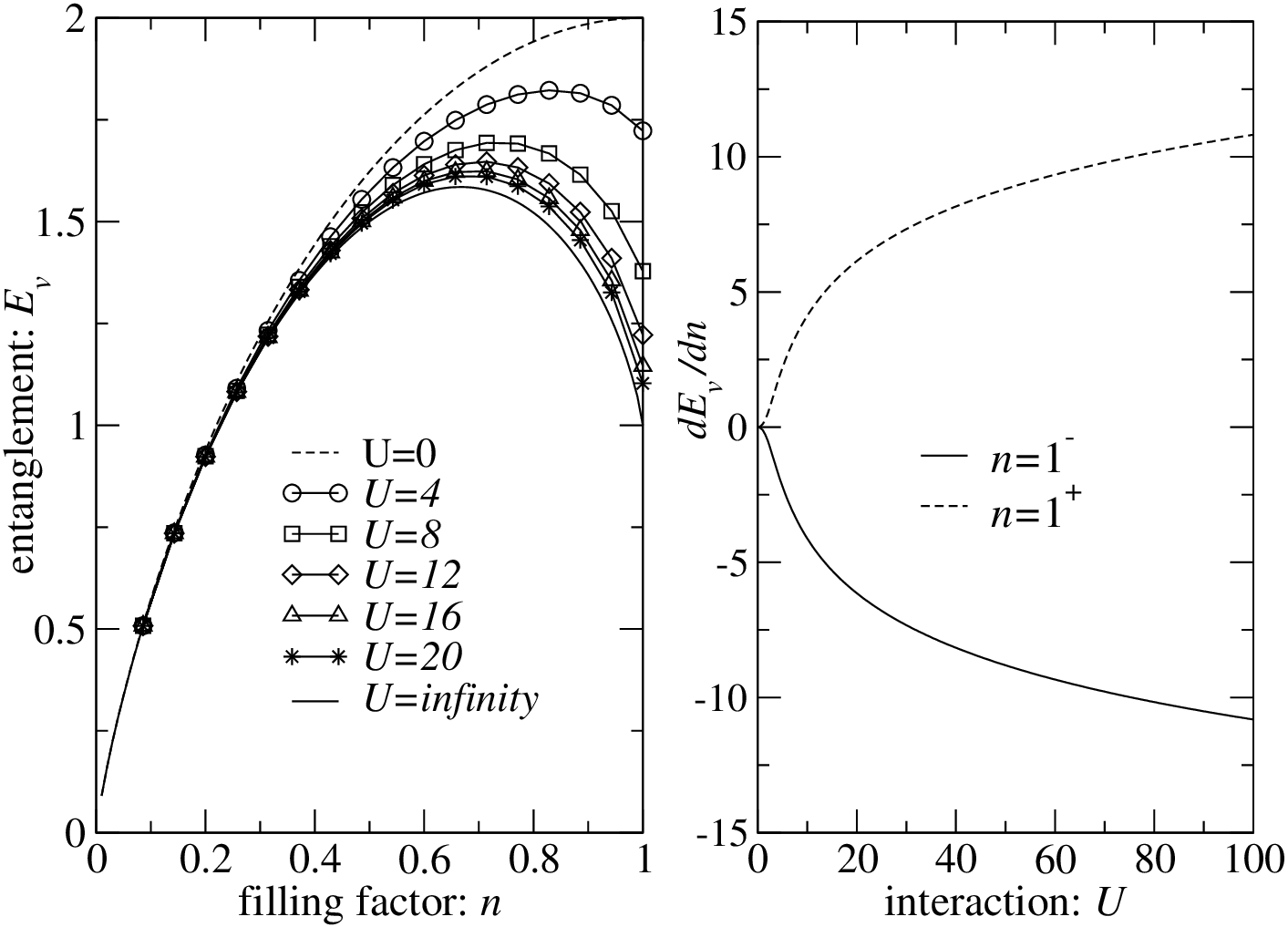}
\caption{\label{figure_lent} The left is the plot of local
entanglement $E_v$ versus the filling factor $n$ for various
interaction strength; the right is the relation between $dE_v/dn$
and $U$ closing to half-filling.}
\end{figure}

{\it The variation of the local entanglement caused by chemical
potential}: By adding a chemical potential term $- \mu \sum_i n_i$
to the Hamiltonian Eq. (\ref{eq:Hamiltonian}), the particle number
of the ground state, hence the filling factor, could be tuned. We
show the relations between local entanglement and the filling
factor $n$ for various on-site couplings $U$ in Fig.
\ref{figure_lent}. We only need to plot the part for $n=N/L<1$
because the part for $n>1$ could simply be obtained by the mirror
image relation, as easily seen from Eq. (\ref{eq:nUtran}), namely
\begin{eqnarray}\label{eq:kgjdflgdg}
E_v(n)=E_v(2-n).
\end{eqnarray}
Fig. \ref{figure_lent} manifests that the ground state of the
half-filled band is not maximally entangled as long as $U>0$,
whereas, the maximum of $E_v$ lies in between $n=2/3$ and $n=1$.
There is no double occupancy when $U=\infty$, which implies that
$w=0$ and $u^+=u^-=N/2L$. Hence we have an analytical expression
of the local entanglement $E_v=-(1-n)\log_2(1-n)-n\log_2(n/2)$
which has a maximum at $n=2/3$. It is worthwhile to point out that
at $1/3$ filling (i.e., $n=2/3$) when $U=\infty$, the ground state
is a singlet of SU$(2|1)$ Lie supersymmetry algebra which
possesses the maximal symmetry allowed, while at 1/2 filling, it
is a SU$(2)$ singlet. For $U=0$ the ground state is invariant
under SO$(4)$ rotation at 1/2 filling. This demonstrates that the
local entanglement reaches a maximum value at the state with
maximal symmetry. Accordingly, the maximum position for
$0<U<\infty$ is expected to lie between $n=2/3$ and $n=1$, which
is confirmed in Fig. \ref{figure_lent}.

Except at half-filling where it becomes a Mott-insulator, the
system is an ideal conductor \cite{EHLieb68,MTakahashib}.
Consequently, the local entanglement $E_v$ is not smoothly
continuous at $n=1$ for $U\neq 0$. It is useful to observe the
derivative of $E_v$ with respect to $U$.
\begin{eqnarray}
\left.\frac{dE_v}{dn}\right|_{n=1^-}=-\left.(\log_2u^+-\log_2z)\left[\frac{1}{2}+2\frac{d\Delta
E}{dU}\right]\right|_{n=1}
\end{eqnarray}
where $\Delta E$ is the gap of charge excitation.
Eq.(\ref{eq:kgjdflgdg}) gives rise to
$\left.dE_v/dn\right|_{n=1^+}=-\left.dE_v/dn\right|_{n=1^-}$.
Obviously, there exists a jump in the derivative of $E_v$  across
the point of insulating phase (see Fig. \ref{figure_lent} right)
unless $U=0$.

\begin{figure}
\includegraphics[width=7cm]{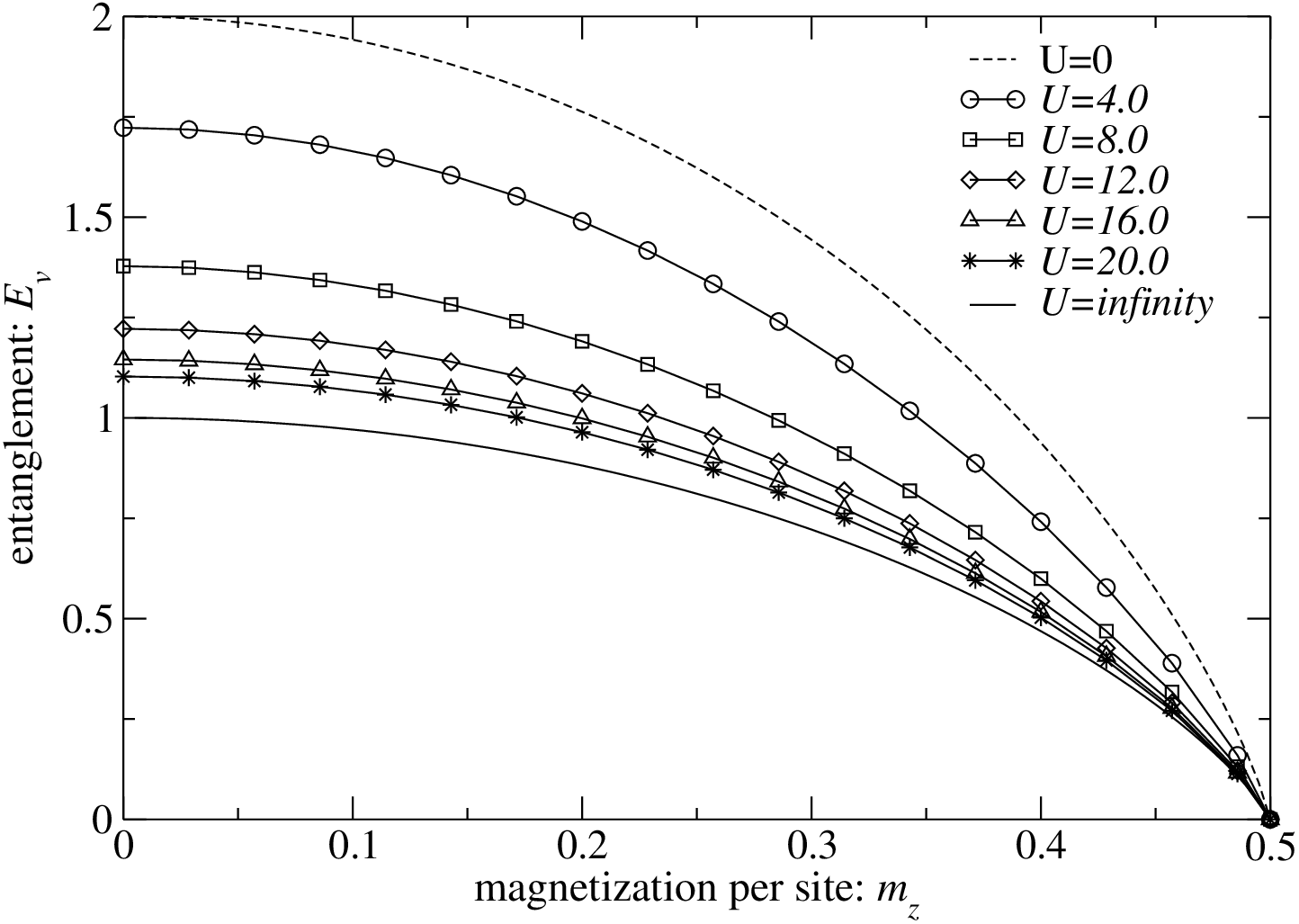}
\caption{\label{figure_ment} The local entanglement $E_v$ versus
the magnetization $m_z=M_z/L$ for different on-site interactions.}
\end{figure}

{\it The dilapidation of the local entanglement induced by the
magnetic field:} Since the Zeemann term commutates with the
original Hamiltonian (\ref{eq:Hamiltonian}), the Bethe-ansatz
solution is still applicable. The application of an external
magnetic field attempts to flip electron spins align along the
magnetic field. There will be more electrons with up spins  than
that with down spins in the presence of magnetic field,
consequently, $u^+ > u^-$. The spin flipping will also diminish
the density of double occupancy, which makes both $w$ and $z$
decrease. As a result, the local entanglement is a descending
function of the magnetization $m_z$ per length. We plot curves of
$E_v$ versus $m_z$ for different interaction strength in Fig.
\ref{figure_ment}. The local entanglement descends to zero when
the magnetization is saturated to $m_z=1/2$. Obviously, the
strength of the on-site interaction suppresses the local
entanglement also in the presence of magnetic field.

>From the above discussions, we find that the local entanglement
manifests distinct features at the point where quantum phase
transition undergoes. The variation of it caused by the chemical
potential and its descendence induced by magnetic field are both
not continuous at the quantum phase transition points. This is
similar to other studies, e.g., the one-dimensional XY model in a
transverse magnetic field \cite{AOsterloh2002,TJOsbornee}, where
the derivative of the pairwise concurrence $C$ with respect to the
dimensionalless coupling constant develops a cusp at the quantum
phase transition point. However, such discontinuity is not
universal. The local entanglement in the one-dimensional Hubbard
model (for fixed chemical potential without external magnetic
field) is smoothly continuous at the transition point driven by
the on-site interaction aforementioned. Similarly, we also
obtained in our recent work on the one-dimensional XXZ mode
\cite{SJGuup}, near the quantum phase transition point $\Delta=1$
where Mott-insulator phase transition occurs, the pairwise
concurrence $C$ is $C=C_0-C_1(\Delta-1)^2$. No discontinuity
occurs in these two cases.

It is indicated \cite{GSTian} recently that two mechanisms may
bring about quantum phase transitions in one-dimensional system.
One is caused by the level crossing of the ground state and the
other arises from the level crossing of the low-lying excited
state where no level crossing occurs at the ground state. In the
later case, the ground-state wavefunction is smoothly continuous
with respect to the variation of parameters that drive the quantum
phase transition, and hence the entanglement is also smoothly
continuous at the quantum phase transition point. For the former
case, the level crossing of the ground state will clearly cause
the entanglement to be none smoothly continuous at the transition
point. Therefore the continuity properties of the first derivative
of the local entanglement might be an ancillary tool to judge the
mechanism of quantum phase transition proposed in \cite{GSTian}.

In summary, we have extensively studied the local entanglement for
the one-dimensional Hubbard model. At half filling, we obtained
the relationship between the local entanglement and the on-site
Coulomb interaction. Our result indicated that the local
entanglement reaches the maximum value at the critical point where
Mott-insulator transition undergoes, and  that the interacting
strength suppress the local entanglement dramatically. Scaling
behavior closing to the critical point $U=0$ was given as
manifests that the local entanglement is an analytical function of
$U$ at this critical point. The asymptotic behavior of the local
entanglement at the strong coupling limit, $U\rightarrow\infty$,
was also given, Furthermore, we calculated the dependence of the
local entanglement on the filling factor and the magnetization.
The variation of local entanglement caused by the chemical
potential shows that the local entanglement reaches maximum at
filling factor $n$ between 2/3 and 1. For any finite on-site
Coulomb interaction $U$, the local entanglement develops a cusp at
$n=1$. At the strong coupling limit, the 1/3 filled band takes the
maximum local entanglement, suggesting that the ground state with
maximal symmetry possesses the maximum magnitude of local
entanglement. The local entanglement decreases with increasing
magnetic field and approaches to zero at saturated magnetization.
For all band fillings, with or without magnetic field on, the
on-site Coulomb interaction always suppresses the local
entanglement.

This work is supported by the NSF China No. 10225419 \& 90103022,
and by the Earmarked Grant for Research from the Research Grants
Council (RGC) of the HKSAR, China (Project CUHK 4037/02P).

\end{document}